\documentclass{jpp}
\usepackage{graphicx}
\usepackage{epstopdf, epsfig}

 \usepackage{amssymb}
 \usepackage{bigints}
 \usepackage{amsmath}
 \usepackage{empheq}
 \usepackage{framed}
 \usepackage{color}
 \usepackage{enumerate}
 \usepackage{ulem}
 \usepackage{bm}
 \usepackage{appendix}
 \usepackage{multicol}
 \numberwithin{equation}{section}
 \usepackage{upgreek}
 \usepackage{textgreek}
\usepackage{color}
\usepackage{scalerel}
\usepackage{stackengine,wasysym}
\usepackage{stmaryrd} %llbrackets
\usepackage{xfrac}    % Works better with other fonts

\newcommand{\be}{\begin{equation}}
\newcommand{\ee}{\end{equation}}
\newcommand{\ba}{\[\begin{aligned}}
\newcommand{\ea}{\end{aligned}\]}
\newcommand{\bea}{\begin{eqnarray}}
\newcommand{\eea}{\end{eqnarray}}
\newcommand{\beann}{\begin{eqnarray*}}
\newcommand{\eeann}{\end{eqnarray*}}
\newcommand{\bs}{\begin{split}}
\newcommand{\es}{\end{split}}

\newcommand*{\cA}{\mathcal{A}} % cal

\newcommand*{\cJ}{\mathcal{J}}

\newcommand*{\cT}{\mathcal{T}}

\newcommand*{\B}{\bm{B}}

\newcommand*{\J}{\bm{J}}

\newcommand*{\dt}{\mathrm{d}}

\newcommand*{\dz}{\partial_z}

\newcommand*{\dth}{\partial_\theta}

\newcommand*{\dl}{\bm{\nabla}}
\newcommand*{\del}{\partial}
\newcommand*{\BD}{\bm{B}\cdot\bm{\nabla}}
\newcommand*{\BOD}{\BO\cdot\dl}

\newcommand*{\jump}[1]{\llbracket #1 \rrbracket}

\newcommand*{\lbr}{\left(}
\newcommand*{\rbr}{\right)}

\newcommand{\iotabar}{\mbox{$\iota\!\!$-}}

\newcommand*{\phibar}{\overline{\phi}}

\newcommand*{\alphabar}{\overline{\alpha}}

\newcommand*{\alphaO}{\alpha^{(0)}}

\newcommand*{\phiO}{\phi^{(0)}}
\newcommand*{\BO}{\B^{(0)}}

\newcommand*{\alphaOne}{\alpha^{(1)}}
\newcommand*{\pOne}{p^{(1)}}

\newcommand*{\phibarOne}{\phibar^{(1)}}
\newcommand*{\phibarPOne}{\phibar^{(1)'}}
\newcommand*{\alphabarOne}{\alphabar^{(1)}}
\newcommand*{\alphabarPOne}{\alphabar^{(1)'}}

\newcommand*{\phiOne}{\phi^{(1)}}

\newcommand*{\ROne}{R^{(1)}}

\newcommand*{\alphaTwo}{\alpha^{(2)}}
\newcommand*{\pTwo}{p^{(2)}}

\newcommand*{\phibarTwo}{\phibar^{(2)}}

\newcommand*{\phiTwo}{\phi^{(2)}}

\newcommand*{\RTwo}{R^{(2)}}

\newcommand*{\phiN}{\phi^{(n)}}
\newcommand*{\phiNpOne}{\phi^{(n+1)}}

\newcommand*{\phibarN}{\phibar^{(n)}}
\newcommand*{\phibarNP}{{\phibar^{(n)}}'}
\newcommand*{\phibarNpOne}{\phibar^{(n+1)}}

\newcommand*{\alphaNmOne}{\alpha^{(n-1)}}
\newcommand*{\alphaN}{\alpha^{(n)}}
\newcommand*{\alphaNpOne}{\alpha^{(n+1)}}

\newcommand*{\alphabarNmOne}{\alphabar^{(n-1)}}
\newcommand*{\alphabarNmOneP}{\alphabar^{(n-1)'}}

\newcommand*{\alphabarNpOne}{\alphabar^{(n+1)}}

\newcommand*{\RN}{R^{(n)}}
\newcommand*{\RNpOne}{R^{(n+1)}}

\newcommand*{\supsub}[3]{{#1}_{#2}^{({#3})}}

\newcommand*{\alphabarNV}[1]{\alphabar^{(#1)}}
\newcommand*{\phibarNV}[1]{\phibar^{(#1)}}

\newcommand*{\phiNV}[1]{\phi^{(#1)}}

\newcommand*\reallywidetilde[1]{\ThisStyle{%
  \setbox0=\hbox{$\SavedStyle#1$}%
  \stackengine{-.1\LMpt}{$\SavedStyle#1$}{%
    \stretchto{\scaleto{\SavedStyle\mkern.2mu\sim}{.5467\wd0}}{.7\ht0}%
  }{O}{c}{F}{T}{S}%
}}

\def\test#1{$%
  \reallywidetilde{#1}\,
  \scriptstyle\reallywidetilde{#1}\,
  \scriptscriptstyle\reallywidetilde{#1}
$\par}
\makeatletter
\newsavebox{\@brx}
\newcommand{\llangle}[1][]{\savebox{\@brx}{\(\m@th{#1\langle}\)}%
  \mathopen{\copy\@brx\mkern2mu\kern-0.9\wd\@brx\usebox{\@brx}}}
\newcommand{\rrangle}[1][]{\savebox{\@brx}{\(\m@th{#1\rangle}\)}%
  \mathclose{\copy\@brx\mkern2mu\kern-0.9\wd\@brx\usebox{\@brx}}}
\makeatother

\shorttitle{3D MHD equilibrium in a periodic cylinder} %Title of paper

\title{ Low-shear three-dimensional equilibria in a periodic cylinder}

\shortauthor{Erin Jaquiery, Wrick Sengupta}
\author{Erin Jaquiery\aff{1,2}, Wrick Sengupta\aff{1}}

\affiliation{\aff{1}Courant Institute of Mathematical Sciences, New York University, New York, New York 10012, USA
\aff{2} Greenwich Academy, 200 N Maple Ave, Greenwich, CT 06830, USA
}

\begin{document}

\maketitle

\begin{abstract}
We carry out expansions of non-symmetric toroidal ideal magnetohydrodynamic (MHD) equilibria with nested flux surfaces about a periodic cylinder, in which physical quantities are periodic of period $2\pi$ in the cylindrical angle $\theta$ and $z$.  The cross-section of a flux surface at a constant toroidal angle is assumed to be approximately circular, and data is given on the cylindrical flux surface $r=1$. Furthermore, we assume that the magnetic field lines are closed on the lowest order flux surface, and the magnetic shear is relatively small. We extend earlier work in a flat-torus by Weitzner [Physics of Plasmas 23, 062512 (2016)]and demonstrate that a power series expansion can be carried out to all orders using magnetic flux as an expansion parameter. The cylindrical metric introduces certain new features to the expansions compared to the flat-torus. However, the basic methodology of dealing with resonance singularities remains the same. The results, even though lacking convergence proofs, once again support the possibility of smooth, low-shear non-symmetric toroidal MHD equilibria.
\end{abstract}

\section{Introduction \label{sec:intro}}
Despite the intrinsic design complexities, the stellarator concept has several benefits over tokamaks: steady-state operation being one of them, and therefore it is an important element of magnetic fusion studies. 
The stellarators are carefully designed such that they possess nested magnetic surfaces. Ideal magnetohydrodynamics (MHD) provides the mathematical description of such an equilibrium. However, the existence of smooth nested nonsymmetric toroidal surfaces in equilibrium is yet to be proven \citep{Garabedian2012computational}. Most current works assume that nested flux surfaces exist, and although several state-of-the-art numerical examples exist, e.g., \citep{Garabedian2012computational,betancourt1976equilibrium,hirshman1983steepest}, in reality, no general mathematical proof of the existence is known and remains an issue \cite{grad1967toroidal,helander2014theory}.

To understand the magnetic field line flow we can apply powerful tools from dynamical systems as such flows may be regarded as a Hamiltonian system with $1\sfrac{1}{2}$ degrees of freedom \citep{caryLittlejohn1983Hamiltonian_B}. In the absence of any continuous symmetry, the field-line flow with finite-magnetic shear is in general non-integrable. Integrability of the Hamiltonian system which differs non-perturbatively from an integrable system is a hard problem. However, for Hamiltonians close to integrable ones we can use KAM theorem \citep{lichtenberg1992regular}, provided the magnetic shear is finite and nonzero everywhere. KAM shows that under small generic perturbations, most surfaces with rational rotation transform break up leading to the formation of magnetic islands and stochastic field lines. Only highly irrational surfaces persist, and they typically show self-similar fractal behavior \citep{hudsonKrauss20173D_cont_B,morrison2000magnetic}. MHD equilibrium with such fractal solutions has been recently studied in \cite{krausHudson2017fractal_pressure}. To describe systems where the magnetic shear is weak or zero in some region, mathematical models like non-twist maps \citep{delcastillo1996nontwist_chaos,morrison2000magnetic} have been developed. The classic KAM theorem does not directly apply \citep{arnol1963smalldiv, abdullaev2006construction} to such maps and generalizations of KAM \citep{delsham_delaLlaves2000kAM_Greene, gonzalez_delaLlave2014singularity} must be considered. V.I.Arnold and co-authors \citep{arnol1963smalldiv,arnold_dynamical_iii} referred to the low-shear systems as ``properly degenerate systems". Arnold showed that \citep[theorem 16 and 17 and the remark afterward, p.~187]{arnold_dynamical_iii} for Hamiltonians of $1\sfrac{1}{2}$ and two degrees of freedom, for all initial conditions the values of the ``action" variables remain forever near their initial values. These low-shear systems were shown to be ``more integrable" than the usual perturbed system in the sense that the measure of the set of tori that disappear under perturbation is exponentially small $O (\exp{(-\text{const}/\epsilon)}$, instead of $O(\sqrt{\epsilon})$ in the non-degenerate case, while the deviation of a perturbed torus from the unperturbed one is of $O(\epsilon)$. The degeneracy condition (low-shear) can be further relaxed \citep{pyartli1969diophantine,sevryuk1995kam,chierchia1982smooth} and so long as magnetic shear is small but non-zero, invariant tori continue to exist.

In the plasma literature, it is known that the behavior of low-shear magnetic field system can be markedly different in the neighborhood of closed field lines (i.e., rational surfaces when they persist) or generic ergodic surfaces \citep{firpo2011study}. Numerous experimental results from Wendelstein VII-A/AS \citep{hirsch2008majorW7AS,brakel2002energytransp_rational_iota_W7AS,brakel1997W7AS_EB_shear} and numerical results \citep{wobig1987localized_pert_w7A} clearly support the idea that optimum confinement is usually found close to certain low-order rational surfaces. For arbitrary perturbations, the islands on these surfaces are
found \citep{wobig1987localized_pert_w7A} to be exponentially small in size in accordance with Arnold's theorem on ``properly degenerate systems." Several authors \citep{grad1973magnetofluid,strauss1981limiting_beta_vacuum} have highlighted the stability of low-shear systems compared to the high shear systems.

Insights from the field line flow make it clear that if we seek nested surfaces everywhere by perturbing an integrable system, the perturbation cannot be generic. In particular, if we look for continuously deformable smooth and continuous non-symmetric solutions of ideal MHD, the pressure and rotation transform profiles cannot be arbitrarily chosen to avoid the singular divisors on rational surfaces \citep{grad1967toroidal,newcomb1959magnetic,hudsonKrauss20173D_cont_B}. However, as discussed earlier, results from dynamical systems, experiments and numerics suggest that low-shear systems in the vicinity of low-order rational surfaces could support such constrained smooth MHD equilibria. Previous work by Weitzner and Sengupta \citep{weitzner2014ideal,weitzner2016expansions,sengupta_weitzner2018vacuum} suggest that in low-shear systems, ideal MHD equilibrium formal expansions can be carried out to all orders around low order rational surfaces with closed field lines. The low-shear closed field line systems provide unique flexibility in avoiding the singular divisor problem as will be shown in details later. This work is, therefore, another step toward supporting the hypothesis that large classes of nested surfaces with smooth pressure and rotation transform profiles might indeed exist.

In this problem, we assume that there is a doubly-periodic cylindrical magnetic flux surface to the lowest order. This approach is an extension of \citep{weitzner2016expansions} calculation done with cartesian geometry. However, rather than using cartesian geometry here, cylindrical geometry is used to address the issue of realistic geometries. The double periodicity of the domain introduces complications in the analysis since they impose stringent restrictions on the structure of the solutions. We have dealt with these constraints following Weitzner’s 2016 approach of carefully eliminating magnetic resonances at each order of the expansion. However, because of the intrinsic cylindrical nature of the domain, specific cylindrical geometry related terms absent in Weitzner 2016 was found. This work concludes that it is possible to carry out a complete power series expansion of the solutions to ideal MHD equations to all orders, by systematically eliminating magnetic resonance at each order.  Our formal procedure indicates that a dense set of nested surfaces can exist provided certain necessary conditions are satisfied. We have not attempted to prove convergence of the series.
    
The paper is structured as follows. In section \ref{sec:basic}, we discuss the basic mathematical structure of three-dimensional MHD equilibrium and the periodicity constraints that come in a toroidal equilibrium. In section \ref{sec:series} and \ref{sec:induct}, we present our main results. We construct ideal MHD equilibrium expansions in the periodic cylindrical domain and discuss the effects of cylindrical metric on the equilibrium. We then use mathematical induction to extend the series expansion to all orders. Finally, we discuss our findings and conclude in section \ref{sec:conclusions}.

 \section{Basic equations}
 \label{sec:basic}
In the following, we set up the formalism for the expansion in the flux coordinate which is a measure of the distance from the lowest order magnetic flux surface where data is given. \cite{weitzner2016expansions} showed that a wide class of such expansions are possible for an MHD equilibrium in a flat topological torus. In the limit of zero pressure gradient, these results include force-free fields. The vacuum field case was later analyzed in \citep{sengupta_weitzner2018vacuum}. The flux surface there was the plane $x=0$, and a particularly simple structure of the field on the surface was assumed. We explore the analogous case here assuming the lowest order flux surface on which data is given to be a cylinder of unit radius. We follow the notation of \citep{sengupta_weitzner2018vacuum} instead of \citep{weitzner2016expansions}.

When the magnetic field $\B$ and the associated current $\J$, both lie on a flux surface $\psi(x,y,z)=$ constant, then both have Clebsch representation of the form,
  \begin{align}
  \B &= \dl \psi \times \dl \alpha \label{a1}\\
  \J &= \dl \zeta \times \dl \psi\equiv \dl \times (\zeta \dl\psi).
    \label{currentJ}
  \end{align}
Since the current $\J$  obtained from curl of $\B$, equation (\ref{currentJ}) suggests that $\B$ can also be expressed as
\begin{align}
\B =\dl\phi +\zeta \dl \psi
\label{BgradPhiform}
\end{align}
 Note that $\phi$ and $\alpha$ can be multivalued. Let us now introduce the toroidal angle coordinates $(\theta,\varphi)$. We obtain easily the Jacobian $\cJ$ and the operator $\BD$ in this coordinate system
 \begin{subequations}
 \begin{align}
 \cJ=\dl \psi \cdot \dl \theta \times  \dl \varphi = \frac{\del(\psi,\theta,\varphi)}{\del(x,y,z)}
  \label{a2}\\
  \BD= \cJ \lbr \alpha_{,\theta} \del_\varphi -\alpha_{,\varphi} \del_\theta\rbr .
   \label{B.D} 
 \end{align}
 \end{subequations}
Here and elsewhere, we use the notation $\alpha_{,\theta}$ to denote partial derivative of $\alpha$ with respect to $\theta$ at fixed $\psi$ and $\varphi$. 
Equating the two different forms of $\B$ given by (\ref{a1}) and (\ref{BgradPhiform}), we get $$(\phi_{,\psi}+\zeta)\dl \psi +\phi_{,\theta}\dl \theta+\phi_{,\varphi}\dl \varphi= \alpha_{,\theta} \dl \psi\times \dl \theta +\alpha_{,\varphi} \dl \psi\times \dl \varphi. $$ Dotting with $ (\dl\varphi\times\dl\psi),(\dl\psi\times\dl\theta)$ and $(\dl \theta\times \dl  \varphi ), $ we obtain the following
 \begin{subequations}
 \begin{align}
 \cJ^{-1}\begin{pmatrix}
\phi_{,\theta}\\ \phi_{,\varphi}
\end{pmatrix}&=
\begin{pmatrix}
g_{\theta\theta} \quad  \quad g_{\theta\varphi}\\ 
g_{\theta\varphi} \quad  \quad g_{\varphi\varphi}
\end{pmatrix}
\begin{pmatrix}
-\alpha_{,\varphi}\\ \alpha_{,\theta}
\end{pmatrix}
\label{a5}\\
\cJ^{-1}\lbr \phi_{,\psi}+\zeta\rbr&= \alpha_{,\theta}\:g_{\varphi \psi}-\alpha_{,\varphi\:}g_{\theta \psi}
 \label{a3}
 \end{align}
 \label{sysa3a4a5}
 \end{subequations}
where $g_{\theta\psi}$ denotes the usual metric coefficient $(\cJ^{-2})\dl\varphi \times \dl \psi \cdot \dl \theta \times \dl \varphi$ etc. 

Next, from the force balance condition ($\J\times\B=p'(\psi)\dl\psi$) we obtain 
\begin{align}
\BD \zeta = p'(\psi).
\label{MDEzeta}
\end{align}
Using (\ref{a3},\ref{MDEzeta}), we finally obtain the generalized Grad-Shafranov equation
\begin{align}
\lbr \alpha_{,\theta} \del_\varphi -\alpha_{,\varphi} \del_\theta\rbr\lbr \phi_{,\psi}-\cJ\lbr \alpha_{,\theta}\:g_{\varphi \psi}-\alpha_{,\varphi\:}g_{\theta \psi}
 \rbr\rbr + \cJ^{-1}p'(\psi)=0
 \label{genGS}
\end{align}
 A magnetic field satisfying ideal MHD equilibrum condition with nested flux surfaces is therefore  characterized by the system (\ref{a5},\ref{genGS}). We now specialize to the straight torus ( a periodic cylinder) in three dimensions in which we take the cylindrical angle $\theta$ and $z$ as the two angles and assume that physical quantities are $2\pi$ periodic. Instead of the cylindrical radius $r$, we shall use $R=r^2/2$. Expressing the various metric coefficients in terms of the cylindrical coordinate system, and using $\cJ =\psi_{,R}$, we obtain
  \begin{subequations}
 \begin{align}
\psi_{,R}\begin{pmatrix}
\phi_{,\theta}\\ \phi_{,z}
\end{pmatrix}=
\begin{pmatrix}
(2R\:\psi_{,R}^2+\psi_{,\theta}^2/2R) \quad  \quad \psi_{,\theta}\psi_{,z}/2R \quad\\ 
\quad \psi_{,\theta}\psi_{,z}/2R \quad\quad  \quad (\psi_{,R}^2+\psi_{,z}^2/2R)
\end{pmatrix}
\begin{pmatrix}
-\alpha_{,z}\\ \alpha_{,y}
\end{pmatrix}
\label{a8}\\
\lbr \alpha_{,\theta} \del_z -\alpha_{,z} \del_\theta\rbr\lbr \phi_{,\psi}+\cJ^{-1}\lbr \alpha_{,\theta}\:\psi_{,z}-\alpha_{,z\:}\psi_{,\theta}
 \rbr/2R\rbr + \cJ^{-1}p'(\psi)=0
\label{a9}.
\end{align}
\label{sysa7a8a9}
\end{subequations}
 Finally, it is convenient to use the inverse representation (details given in \cite{weitzner2016expansions,sengupta_weitzner2018}) and assume $R=R(\psi,\theta,z)$ rather than $\psi=\psi(R,\theta,z)$
\begin{subequations}
 \begin{align}
R_{,\psi}\begin{pmatrix}
\phi_{,\theta}\\ \phi_{,z}
\end{pmatrix}=
\begin{pmatrix}
(2R+R_{,\theta}^2/2R) \quad  \quad R_{,\theta}R_{,z}/2R\\ 
\quad R_{,\theta}R_{,z}/2R \quad  \quad (1+R_{,z}^2/2R)
\end{pmatrix}
\begin{pmatrix}
-\alpha_{,z}\\ \alpha_{,\theta}
\end{pmatrix}
\label{a10a11}\\
\lbr \alpha_{,\theta}\dz- \alpha_{,z}\dth \rbr\lbr\phi_{,\psi}-(R_{,z}\alpha_{,\theta}-R_{,\theta}\alpha_{,z})/2R \rbr +p'(\psi) R_{,\psi}=0
\label{a12}
\end{align}
\label{sysa10a11a12}
\end{subequations}
In the following, we shall work mostly with equation system (\ref{sysa10a11a12}). Equation (\ref{a12}) is an inhomogeneous magnetic differential equation and as shown by \citep{newcomb1959magnetic} there are stringent restrictions imposed on solutions of such equations. Before presenting details of the expansion, we shall briefly discuss the basic structure of the periodicity constraints 
\subsection{Magnetic differential equations and consistency conditions}
The magnetic differential operator $(\BD)$ in the inverse representation is given by
\begin{align}
\BD=\frac{1}{R_{,\psi}}\lbr \alpha_{,\theta}\dz- \alpha_{,z}\dth \rbr\:.
\label{BDinvI}
\end{align}
In the analysis of the ideal MHD system we often encounter the so called the magnetic differential equation (MDE), which is of the form $\BOD f=Q$. We assume $\BO$ is a closed magnetic field line and 
$Q$ is an arbitrary function of all the three variables. We can rewrite the MDE as follows 
\begin{align}
\lbr \alphaO_{,\theta}\dz- \alphaO_{,z}\dth \rbr f=G.
\label{MDE1}
\end{align}
The goal is to solve for $f$ in equation (\ref{MDE1}) subject to the periodicity constraint that arises due to the toroidal nature of the magnetic surface. The requirement that physical variables be single-valued restricts $f$ to the form (cf. \citep{weitzner2016expansions})
\begin{align}
f=g(\psi) \theta + h(\psi) z +\tilde{f}(\psi,\theta,z)
\label{fform}
\end{align}
where $g$ and $h$ are functions of $\psi$ and $\tilde{f}$ is a periodic function of the angles. If $f$ is a single-valued function then $(g,h)$ are identically zero. Integrating along the closed field line, we obtain the following periodicity constraint 
\begin{align}
\oint_{\alphaO} \frac{d\theta}{-\alphaO_{,z}} G = \oint_{\alphaO} \frac{dz}{\alphaO_{,\theta}}G =\overline{G}(\psi).
\label{Newcombcondition}
\end{align}
The  subscript $\alphaO$ indicates that the integration is carried out along a constant field line label $\alphaO$, and $\overline{G}$ is a function of $\psi$ alone. This condition has to be satisfied in order to avoid magnetic resonance and obtain a self consistent solution. To see the connection between the jump of the multivalued function after a complete circuit along the closed magnetic field, and $\overline{G}$, we define the jump in $f$ as
\begin{align}
\llbracket f \rrbracket \equiv \oint_{\alphaO} \frac{dz}{\alphaO_{,\theta}} \lbr \alphaO_{,\theta}\dz- \alphaO_{,z}\dth \rbr f.
\label{jump}
\end{align}
If the closed field line closes on itself after $m$ ``toroidal'' turns in $z$ and $n$ ``poloidal'' turns in $\theta$ then the jump in $f$ is given by $\jump{f}= m h(\psi) - n g(\psi) =\overline{G}(\psi) $. Therefore, the consistency condition \eqref{Newcombcondition} implies that the jump must be independent of the field line and must be only a function of the flux-surface $\psi$. 

Provided the consistency condition \eqref{Newcombcondition} is satisfied, we can solve for $f$ up to an undetermined function $\bar{f}(\psi,\alphaO)$  which is a homogeneous solution of the MDE i.e., $\BOD \bar{f} =0$. This is an important aspect of the problem because this allows a degree of freedom ($\bar{f}(\psi,\alphaO)$) that we can utilize to satisfy the periodicity constraint of another MDE. The $\alphaO$ dependence of $\bar{f}$ is possible only for closed field line systems. We shall call the homogeneous solution $\bar{f}(\psi,\alphaO)$ a free-function that can be utilized to ensure solvability of another MDE.
\section{Series expansion in $\psi$ about a given flux surface}
\label{sec:series}
 We shall now use the system (\ref{a10a11},\ref{a12})to explore the possibility of a formal power series expansion of $(R(\psi,y,z),\phi(\psi,y,z),\alpha(\psi,y,z))$ in the coordinate $\psi$. The expansion repeats the process laid out in \citep{weitzner2016expansions}. We expand as 
 \begin{align}
 (R,\phi,\alpha)=\sum_{n\geq 0}(R^{(n)}(y,z),\phi^{(n)}(y,z),\alpha^{(n)}(y,z)) \psi^n 
 \label{a13}
 \end{align}
 We assume that $R=1/2$ is a magnetic surface with circular cross-section so that $R^{(0)}=1/2$. Furthermore, the multivalued functions $(\phi^{(n)},\alpha^{(n)})$ must be of the form  given by equation (\ref{fform}) with constant $g^{(k)}$ and $h^{(k)}$. 
 
We now rewrite the system (\ref{sysa10a11a12}) in a form that is more amenable to expansions
\begin{subequations}
\begin{align}
R_{,\psi}\phi_{,\theta}+\alpha_{,z}&= -2\lbr R-\frac{1}{2} \rbr\alpha_{,z} +R_{,\theta}\cT \label{T1}\\
R_{,\psi}\phi_{,z}-\alpha_{,\theta}&= R_{,z}\cT
\label{T2}\\
\lbr \alpha_{,\theta}\dz -\alpha_{,z} \partial_{\theta} \rbr (\phi_{,\psi}-\cT) &= -p' R_{,\psi}
\label{T3}\\
\text{where,}\quad \cT &= \frac{1}{2R}(R_{,z}\alpha_{,\theta}-R_{,\theta}\alpha_{,z}).
\end{align}
\label{Tform}
\end{subequations}
 We note that the terms $\cT,(R-1/2)$ are of at least first order in $\psi$ while the terms $(R_{,\theta},R_{,z})\cT$ are at least second order.
 We find easily to lowest order that
 \begin{subequations}
 \begin{align}
 R^{(1)}\phi^{(0)}_{,\theta} +\alpha^{(0)}_{,z}&=0 \label{a14a}\\
 R^{(1)}\phi^{(0)}_{,z}  -\alpha^{(0)}_{,\theta}&=0\label{a14b}\\
 \lbr \alpha^{(0)}_{,\theta}\dz- \alpha^{(0)}_{,z}\dth \rbr\phi^{(1)}&=-\pOne \ROne \label{a14c}
 \end{align}
 \label{ZeroO}
 \end{subequations}
 From (\ref{a14a},\ref{a14b}) we note that $(\phiO,\alphaO)$ are orthogonal. Assuming the lowest order field lines close on themselves after $m$ toroidal and $n$ poloidal turns, from (\ref{a14c}) we find that $\ROne$ must satisfy
 \begin{align}
 \oint_{\alphaO} \frac{dz}{\alphaO_{,\theta}} \ROne = -\oint_{\alphaO} \frac{d\theta}{\alphaO_{,z}} \ROne=\text{constant independent of $\alphaO$.}
 \label{NewcombR1}
 \end{align}
Once we have found functions $(\phiO,\alphaO,\ROne)$ satisfying system(\ref{ZeroO}) and (\ref{NewcombR1}), we can solve for $\phiOne$ up to an as yet undetermined homogeneous solution $\phibarOne=\phibarOne(\alphaO)$
\begin{align}
\phiOne= \phibar^{(1)}(\alphaO) -\pOne \int_{\alphaO}\frac{dz}{\alphaO_{,\theta}}\ROne.
\label{phi1sol}
\end{align}
To first order, we have
\begin{subequations}
 \begin{align}
 2 R^{(2)}\phi^{(0)}_{,\theta}  +\alpha^{(1)}_{,z} =&-\ROne \lbr \phiOne_{,\theta}+2 \alphaO_{,z} \rbr\label{a15a}\\
 2 R^{(2)}\phi^{(0)}_{,z}  -\alpha^{(1)}_{,\theta}=& -\ROne \phiOne_{,z}\label{a15b}\\
 \lbr \alpha^{(0)}_{,\theta}\dz- \alpha^{(0)}_{,z}\dth \rbr \lbr 2 \phi^{(2)}-\lbr \ROne_{,z}\alphaO_{,\theta}  -\ROne_{,\theta}\alphaO_{,z}\rbr \rbr =& -\lbr \alpha^{(1)}_{,\theta}\dz  -\alpha^{(1)}_{,z}\dth \rbr\phiOne \nonumber\\
 &-\pTwo \ROne -2\pOne \RTwo\:. \label{a15c}
 \end{align}
  \label{1stO}
 \end{subequations}
 Eliminating $\RTwo$ between (\ref{a15a},\ref{a15b}), we obtain a magnetic differential equation for $\alphaOne$
 \begin{align}
  \alpha^{(0)}_{,\theta}\alpha^{(1)}_{,z}-\alpha^{(0)}_{,z}\alpha^{(1)}_{,\theta}= -\ROne \lbr \alphaO_{,\theta}\phiOne_{,\theta}+ \alphaO_{,z}\phiOne_{,z}\rbr -2\ROne \alphaO_{,\theta}\alphaO_{,z}.
 \label{MDEalpha1}
 \end{align}
 This equation looks almost identical to the corresponding $\alphaOne$ equation (28) in the cartesian geometry (Weitzner 2016) except for the last term. This term appears solely because of the cylindrical geometry of the lowest order flux surface.
Clearly, because there is an MDE involved in the solution for $\alphaOne$, there must be a check for consistency before solving for the next order solution. The consistency condition is
\begin{align}
-\oint_{\alphaO} \frac{dz}{\alphaO_{,\theta}}\ROne \lbr \alphaO_{,\theta}\phiOne_{,\theta}+ \alphaO_{,z}\phiOne_{,z}\rbr +2\oint_{\alphaO} dz \ROne \alphaO_{,z} =\text{constant}
\label{NewcombO2}
\end{align}
 We note that the second integral in (\ref{NewcombO2}) is absent in a planar geometry, and the free-function $\phibar^{(1)}(\alphaO) $ appears in (\ref{NewcombO2}) through the term
\begin{align}
-{\phibarPOne}\oint_{\alphaO} \frac{dz}{\alphaO_{,\theta}}\ROne \lbr\lbr \alphaO_{,\theta}\rbr^2+\lbr \alphaO_{,z}\rbr^2\rbr \label{phibar1contrib}
\end{align} 
 We now show that the left side of the above equation can be made constant through a suitable choice of $\phi^{(1)'}$. We note that $\ROne$ and $\alphaO$ are periodic functions of $(\theta,z)$ and the integration is being done along a constant $\alphaO$ line. Therefore, using implicit function theorem we can rewrite any $z$ dependence in terms of
$(\theta,\alphaO)$ dependence. After the loop integration is done from $\theta=0$ to $\theta= 2 m \pi$, the result can only be a periodic function of $\alphaO$ since all the terms in the integrand involve single valued and periodic functions of the angles. Provided the integral in (\ref{phibar1contrib}) does not vanish, we can choose $\phibar^{(1)}(\alphaO) $ so that the left hand side is a constant. Thus the consistency condition holds. 

We solve the MDE (\ref{MDEalpha1}) for $\alphaOne$ as
\begin{align}
\alphaOne=\alphabarOne(\alphaO) -\int_{\alphaO} \frac{dz}{\alphaO_{,\theta}}\ROne \lbr \alphaO_{,\theta}\phiOne_{,z}+ \alphaO_{,z}\phiOne_{,\theta}\rbr +2\int_{\alphaO} dz \ROne \alphaO_{,z}
\label{alpha1sol}
\end{align}
where $\alphabarOne(\alphaO)$ is a homogeneous solution to MDE periodic in $\alphaO$. 
Next we turn to $\RTwo$. In order to solve for $\RTwo$, either of equations (\ref{a15a},\ref{a15b}) can be used together with (\ref{alpha1sol}). The contribution of the free-function $\alphabarOne$ to $\RTwo$ can be easily shown to be
\begin{align}
2\RTwo=\ROne \alphabarPOne(\alphaO)+...
\label{R2D2}
\end{align}

The final quantity to be determined in this order is $\phiTwo$.  In equation (\ref{a15c}), the MDE operator acts on $\phiTwo$ with  all other quantities known. We can substitute the expressions of $(\alpha^{(1)}_{,\theta}$ and $\alpha^{(1)}_{,z})$ from (\ref{a15a},\ref{a15b}) in (\ref{a15c}) to obtain 
\begin{align}
\lbr \alpha^{(0)}_{,\theta}\dz- \alpha^{(0)}_{,z}\dth \rbr \lbr 2 \phi^{(2)}-\lbr \ROne_{,z}\alphaO_{,\theta}  -\ROne_{,\theta}\alphaO_{,z}\rbr \rbr = \nonumber\\
 - \ROne \lbr \lbr\phiOne_{,\theta}\rbr^2+\lbr\phiOne_{,z}\rbr^2 +2 \phiOne_{,\theta}  \alphaO_{,z} +\pTwo\rbr.
\label{MDEphi1sqr}
\end{align}
Note that the $\RTwo$ term in (\ref{a15c}) drops out completely in (\ref{MDEphi1sqr}). Therefore, the free-function in (\ref{R2D2}) can not be used to satisfy the consistency condition of the MDE (\ref{MDEphi1sqr}). The periodicity constraint is therefore,
\begin{align}
\oint_{\alphaO} \frac{dz}{\alphaO_{,\theta}}\ROne  \lbr \lbr\phiOne_{,\theta}\rbr^2+\lbr\phiOne_{,z}\rbr^2 +2 \phiOne_{,\theta}  \alphaO_{,z} \rbr = \text{constant}.
\label{Newcombphi2}
\end{align}
Thus, we have three integral constraints, (\ref{NewcombR1}),(\ref{NewcombO2}) and (\ref{Newcombphi2}) that must be satisfied by the first order quantities $\ROne$ and $\phiOne$. The only free-function available is $\phibarOne(\alphaO)$. Provided we can satisfy all the three constraints, the solution for $\phiTwo$ then takes the form
\begin{align}
\phiTwo=\phibarTwo(\alphaO)-&\frac{1}{2}\lbr \int_{\alphaO} \frac{dz}{\alphaO_{,\theta}}\lbr \alpha^{(1)}_{,\theta}\dz -\alpha^{(1)}_{,z}\dth \rbr\phiOne \right.\nonumber\\
&+\left.\pTwo \int_{\alphaO} \frac{dz}{\alphaO_{,\theta}}\ROne +2\pOne \int_{\alphaO} \frac{dz}{\alphaO_{,\theta}}\RTwo \rbr
\label{phi2sol}
\end{align}
where $\phibarTwo$ is once again a free-function that is undetermined at this order but will be fixed when we solve MDE for $\alphaTwo$ at next order. At this point in the calculations, the functions $(\alphaOne,\phiTwo, 2\RTwo)$ are determined up to $(\alphabarOne, \phibarNV{2},\ROne \alphabarPOne)$. 

In Appendix \ref{App:AppendixA}, we show that the lowest order solutions of the system (\ref{ZeroO}) in a flat torus \citep{weitzner2016expansions} is still valid for a periodic cylinder provided a discrete symmetry called stellarator symmetry \citep{dewar1998stellarator} is invoked. We obtain an explicit solution of system \eqref{ZeroO} that satisfy all the three integral constraints (\ref{NewcombR1}),(\ref{NewcombO2}) and (\ref{Newcombphi2}).

\section{Inductive proof to carry out the expansion to all orders}
\label{sec:induct}
We now show a recurring pattern in the construction of the power series expansions of the functions $(\phi,\alpha, R)$. At every order say $O(n)$, the calculation begins with three unknowns $(\phiNV{n+1},\alphaN,\RN)$ being solved for and two free-functions $\phibarN(\alphaO)$ and $\alphabarNV{n-1}(\alphaO)$ available from $O(n-1)$. These free-functions can be used for satisfying the periodicity requirement of $\phiNV{n+1}$ and $\alphaN$. At the end of the calculation, we obtain solutions for all the three unknown functions and two new free-functions $\phibarNpOne(\alphaO),\alphabarNV{n}(\alphaO)$  for later use. We can, therefore, use mathematical induction to show that this expansion can be carried out to all orders. We give a detailed proof of the induction process below.

To $O(n)$, we obtain the following set of equations that determine the unknowns $(\phiNpOne,\alphaN,\RNpOne)$.
\begin{subequations}
 \begin{align}
 (n+1) \RNpOne\phi^{(0)}_{,\theta}+n\RN \phiOne_{,\theta}+\ROne \phiN_{,\theta} +\alpha^{(n)}_{,z}+&\nonumber \\
 2 \ROne \alphaNmOne_{,z} +2 \RN \alphaO_{,z} &= \supsub{F}{1}{n} \label{Ntha}\\
 (n+1) \RNpOne\phi^{(0)}_{,z}  +n\RN \phiOne_{,z}+\ROne \phiN_{,z} -\alpha^{(n)}_{,\theta} &= \supsub{F}{2}{n}
 \label{Nthb}\\
 \lbr \alpha^{(0)}_{,\theta}\dz- \alpha^{(0)}_{,z}\dth \rbr((n+1) \phi^{(n+1)})+\lbr \alpha^{(1)}_{,\theta}\dz- \alpha^{(1)}_{,z}\dth \rbr(n \phi^{(n)}) +&\nonumber\\
 \lbr \alpha^{(n)}_{,\theta}\dz  -\alpha^{(n)}_{,z}\dth \rbr\phiOne
  +n \pTwo \RN+(n+1)\lbr \pOne \RNpOne +p^{(n+1)} \ROne \rbr &=\supsub{F}{3}{n}. \label{Nthc} 
 \end{align}
  \label{Nth}
 \end{subequations}
Here and elsewhere we use the notation $\lbr \supsub{F}{m}{n}, \supsub{\overline{F}}{m}{n}\rbr$ to denote already completely determined functions from the lower orders and their line averages along the closed field line. Eliminating $\RNpOne$ between (\ref{Ntha},\ref{Nthb}) and using \eqref{ZeroO} and \eqref{MDEalpha1}, we obtain a magnetic differential equation for $\alphaN$,
 \begin{align}
  \lbr \alpha^{(0)}_{,\theta}\dz -\alpha^{(0)}_{,z} \dth \rbr \alpha^{(n)}+\ROne \lbr \alphaO_{,\theta}\phiN_{,\theta}+ \alphaO_{,z}\phiN_{,z}\rbr 
  +2\ROne& \alphaNmOne_{,z} \alphaO_{,\theta} \nonumber\\ 
 -\frac{n\RN}{\ROne} \lbr \alpha^{(0)}_{,\theta}\dz -\alpha^{(0)}_{,z} \dth \rbr \alpha^{(1)} &=\supsub{F}{4}{n}. \label{MDEalphaN}
\end{align}
Similarly, eliminating $\alphaN_{,z}$ and $\alphaN_{,\theta}$ in \eqref{Nthc} using \eqref{Ntha},\eqref{Nthb} and \eqref{MDEphi1sqr} we obtain a MDE for $\phiNpOne$, 
\begin{align}
 \lbr \alpha^{(0)}_{,\theta}\dz -\alpha^{(0)}_{,z} \dth \rbr \lbr (n+1)\phiNpOne\rbr
 +\lbr \alpha^{(1)}_{,\theta}\dz -\alpha^{(1)}_{,z} \dth \rbr \lbr n\phiN\rbr 
 +2\ROne &\alphaNmOne_{,z}  \phiOne_{,\theta}\nonumber\\
 +\ROne \lbr \phiN_{,\theta}\phiOne_{,\theta}+\phiN_{,z}\phiOne_{,z} \rbr - \frac{n \RN}{\ROne}\lbr \alpha^{(0)}_{,\theta}\dz -\alpha^{(0)}_{,z} \dth \rbr \lbr 2\phiTwo-\cT^{(1)}\rbr &=\supsub{F}{5}{n},
 \label{MDEphiNp1}
 \end{align}
where $\cT^{(1)}=-\lbr \ROne_{,z}\alphaO_{,\theta}  -\ROne_{,\theta}\alphaO_{,z}\rbr.$

Alternatively, we can nonlinearly eliminate $R_{,\psi}$ between \eqref{T1}, \eqref{T2}, and $\alpha_{\theta},\alpha_{,z}$ from \eqref{T3} using \eqref{T1} and \eqref{T2} to get
\begin{align}
(\alpha_{,z}\phi_{,z}+\alpha_{,\theta}\phi_{,\theta})+2\lbr R-\frac{1}{2}\rbr\alpha_{,z}\phi_{,z}=& \cT (R_{,\theta}\phi_{,z}-R_{,z}\phi_{,\theta})
\label{NLMDE1}\\
R_{,\psi}\frac{\partial}{\del \psi}\lbr p(\psi)+\frac{1}{2}\lbr \lbr{\phiO_{,\theta}}\rbr^2+\lbr{\phiO_{,z}}\rbr^2 \rbr \rbr =& \:(\alpha_{,\theta} \dz -\alpha_{,z}\dth)\cT+\cT (R_{,\theta}\phi_{,z}-R_{,z}\phi_{,\theta})
\label{NLMDE2}
\end{align}
Carrying out the expansions of (\ref{NLMDE1}, \ref{NLMDE2}) to $O(n)$, we recover the MDEs (\ref{MDEalphaN}, \ref{MDEphiNp1}).

We assume that $(\phiN,\alphaNmOne, n\RN)$ are known up to $\phibarN$, $\alphabarNmOne$, and $\ROne \alphabarNmOneP$. $\phibarN$ and $\alphabarNmOne$ are both functions of $\alphaO$. 
Averaging over the field lines, we get the following consistency conditions
\begin{align}
 \cA_2\: \phibarNP -\alphabarNmOneP \cA_1+ \supsub{\overline{F}}{4}{n}(\alphaO) =\text{constant}
\label{NewcombalpN}\\
(n+1)\cA_1\: \phibarNP -2\alphabarNmOneP \cA_3+ \supsub{\overline{F}}{4}{n} (\alphaO)=\text{constant} \nonumber
\end{align}
where
\begin{align}
\cA_1&=\llbracket \alphaOne\rrbracket, \quad \cA_3= \llbracket \phiTwo\rrbracket-\oint_{\alpha^{(0)}}\dt \theta\:\ROne\phiOne_{,\theta} \label{cAs}\\
\cA_2 &= \oint_{\alphaO} \frac{dz}{\alphaO_{,\theta}}\ROne \lbr \lbr \alphaO_{,\theta}\rbr^2+ \lbr \alphaO_{,z}\rbr^2\rbr \nonumber
\end{align}
If the necessary condition for invertibility of the system \eqref{NewcombalpN} 
\begin{align}
\lbr -2\cA_2 \cA_3+(n+1)\cA_1^2\rbr \neq 0
\label{determ}
\end{align}
is satisfied, we can obtain $\phibarNP$ and $\alphabarNmOneP$ from \eqref{NewcombalpN}. In particular, when $\cA_2 \cA_3<0$ and $\cA_1 \neq 0$, \eqref{determ} is satisfied. This allows us to solve the MDEs for $\alphaNpOne$, $\phiNpOne$ and $(n+1)\RNpOne$ up to functions $\alphabarNpOne(\alphaO),\phibarNpOne(\alphaO)$ and  $\ROne \alpha^{(n+1)'}(\alphaO)$ respectively.

Thus, we have outlined a general procedure to satisfy the MHD equilibrium equations to an arbitrary order $(n)$. At each order $O(n)$, we fix two free-functions of $\alphaO$, $\lbr \phibarN \:,\: \alphabarNmOne \rbr$, obtained from the previous order and gain two new free-functions $\lbr \phibarNpOne(\alphaO)\:,\:\alphabarNV{n}(\alphaO) \rbr$. Since self-consistent equilibrium solutions could be found for $n=1$, as shown in Appendix \ref{App:AppendixA}, and for any arbitrary $n$ when the condition \eqref{determ} holds, we prove that solutions can be obtained for all orders by the induction hypothesis.

%Finally, when we have the Taylor series in $(\psi,\theta,z)$ variables, we can invert the series to find the magnetic flux surface $\psi$ in terms of the cylindrical coordinates, i.e., recover $\psi=\psi(R,\theta,z)$. 

\section{Discussion}
\label{sec:conclusions}
In this paper we have extended earlier work (Weitzner 2016) on MHD equilibrium expansions in a flat topological torus to a periodic cylinder, thereby allowing slightly more realistic geometry. We have considered the flux surfaces to have an approximately circular cross-section, so that cylindrical geometry is appropriate. We assume data on the cylindrical flux surface of unit radius, which has closed magnetic field lines. We have shown that a power series expansion in the magnetic flux coordinate can be carried out to all orders by eliminating magnetic resonances at each order. Without discussing the convergence of our series, we have shown that it is possible that a broad class of low-shear smooth non-symmetric MHD toroidal equilibrium exist in a straight torus.

We started with the ideal MHD equations in the inverse formalism, where the flux function and the two cylindrical angles $(\psi,\theta,z)$, are used as coordinates, and $R=r^2/2$ is considered a function of the coordinates $R(\psi,\theta,z)$. We expanded the equations in power series of the magnetic flux $\psi$ about $R=1/2$ (or  $r=1$). Magnetic differential equations (MDEs) were solved self-consistently to find the first order solutions consistent with the periodicity requirements. We found that the cylindrical metric introduces new terms in the expansions which do not appear in the flat torus case analyzed earlier in Weitzner 2016. We have outlined a general procedure to deal with such terms and eliminate magnetic resonances systematically.

Throughout the calculation, we obtain enough such free-functions (functions of the lowest order closed field line label), which provide flexibility in the calculations as they can be chosen judiciously to avoid magnetic resonances. We found a repeating pattern for the higher order equations; at each order, a free-function entered the problem from the previous order, then the free-function was fixed by eliminating the possibility of any magnetic resonance to that order, and later self-consistent solutions of the higher order system were found. Finally, one new free-function emerged as a homogeneous solution to an MDE. This process repeats for the next order. Using mathematical induction, we conclude that the perturbation series can be extended to all orders. We note here that closed field line systems offer this unique flexibility because, in an ergodic field line system, the free-functions can only be constants and therefore, satisfying periodicity constraints like those discussed here would be impossible.

The work presented here relates to low-shear devices like W7-X and HSX in that it supports the ides that low-shear devices may have good surfaces if the boundaries and external fields are chosen carefully. Therefore, this work supports the low shear approach to equilibrium systems. The equations presented here could apply to large aspect ratio stellarators with a low-shear and cross-section that is approximately circular. Stellarators such as W7-X and HSX have low magnetic shear, but their cross-sections are not necessarily circular. A generalization to include non-circularity of cross-section could be carried out in the future. 

We can extend the work in several other directions. Firstly, we would like to carry out an expansion about a generic analytic surface analogous to the cylindrical surface presented here. Furthermore, though the series is shown to extend to all orders, it is still unclear whether these series converge or diverge. It seems reasonable to assume that the series is valid in a small region near $r=1$, but we do not expect it to be convergent near $r=0$ which has a coordinate singularity for the cylindrical geometry. We shall address the issue of convergence in the future.

This work was done as part of The NYU Courant Institute Girls' Science, Technology, Engineering, and Mathematics  (NYU GSTEM) Summer Program. E.J. would like to thank  Harold Weitzner; not only was his prior work the basis of these calculations, but his help and support were fundamental to this project. E.J would also like to thank the NYU GSTEM program for providing this opportunity, especially Catherine Tissot for her organization, Hannah Boland for being such a great tutor, and Monica Parham for the helping with everything along the way.
W.S. acknowledges helpful discussions with Harold Weitzner, Tonatiuh Sanchez-Vizuet and Geoffery McFadden.
This research was partly funded by the US DOE Grant No. DEFG02-86ER53223.

\appendix

\section{Exact solutions of the lowest order quantities} \label{App:AppendixA}
The constraint (\ref{NewcombR1}) suggests that we look for solutions where $\ROne$ is of the form 
\begin{align}
\ROne= -\alphaO_{,\theta}\alphaO_{,z}.
\label{R1form}
\end{align}
Eliminating $\phiO$ from (\ref{a14a},\ref{a14b}) and using (\ref{R1form}), we find a necessary condition
\begin{align}
\alphaO_{,z\theta}=0.
\label{alphaztheta}
\end{align}

A solution of \eqref{alphaztheta} that has the form (\ref{fform}) is  
 \begin{subequations}
 \begin{align}
 \alpha^{(0)}= \mu(y)-\nu(z),\quad  \mu(y)=m y +P(y),\quad \nu(z)=n z +Q(z).
 \label{var_sep}
 \end{align}
 \end{subequations}
 where P and Q are arbitrary $2\pi$ periodic functions in their arguments and the integers $(m,n)$ are relatively prime. It follows that
 \begin{subequations}
 \begin{align}
 R^{(1)}(y,z)&=\mu'(y)\nu'(z)\\
 \phi^{(0)}(y,z)&=\int\frac{dy}{\mu'(y)}+\int\frac{dz}{\nu'(z)}\\
 \phiOne &= \phibarOne(\alphaO) -\pOne \nu(z) \:.
 \label{phi1solexact}
 \end{align}
 \label{exactsol}
 \end{subequations}
We shall now show that the integral constraints (\ref{NewcombO2}) and (\ref{Newcombphi2}), can be satisfied for a system with stellarator symmetry \citep{dewar1998stellarator}. We require that under the transformation  $(z,\theta)\rightarrow (-z,-\theta)$, $\alphaO$ change sign while $\ROne$ does not. From \eqref{var_sep} and \eqref{phi1solexact}, we see that $P,Q$ must be odd functions of their arguments.  We find that the term occurring in \eqref{NewcombO2} due to the cylindrical nature of the geometry, vanishes i.e., 
\begin{align}
\oint_{\alphaO} \dt z\:\ROne \alphaO_{,z}=\oint_{\alphaO} \dt z\:\alphaO_{,\theta}\lbr \alphaO_{,z}\rbr^2 =0.
\label{geometric01}
\end{align}
This follows since under the discrete symmetry, the integrand in \eqref{geometric01} does not change sign while the integral changes sign. 
Similarly, using \eqref{geometric01} and $\phiOne_{,\theta}= \phibarPOne \alphaO_{,\theta}$ that follows from \eqref{phi1solexact}, the geometry related integral in (\ref{Newcombphi2}) simplifies to
\begin{align}
\oint_{\alphaO}\frac{\dt z}{\alphaO_{,\theta}} 2\ROne \phiOne_{,\theta} \alphaO_{,z} =-2\phibarPOne \oint_{\alphaO} \dt z\:\ROne \alphaO_{,z}=0\:.
\label{geometric02}
\end{align}
To calculate the remaining integrals in (\ref{NewcombO2}) and (\ref{Newcombphi2}) we note that
\begin{align}
\oint_{\alphaO}\frac{\dt z}{\alphaO_{,\theta}}\ROne \lbr \alphaO_{,\theta}\rbr^2 = \oint \dt \theta\: \mu'(\theta)^3 = \text{constant} \nonumber\\
\oint_{\alphaO}\frac{\dt z}{\alphaO_{,\theta}}\ROne \lbr \alphaO_{,z}\rbr^2 = \oint \dt z\: \nu'(z)^3 = \text{constant}
\label{quad_integrals}\\
\lbr \lbr\phiOne_{,\theta} \rbr^2+ \lbr\phiOne_{,z} \rbr^2\rbr = \lbr \phibarPOne \rbr^2 \lbr \alphaO_{,\theta}\rbr^2 +\lbr \phibarPOne +\pOne\rbr^2\lbr \alphaO_{,z}\rbr^2  \nonumber
\end{align}
Using \eqref{quad_integrals} we find that all the terms in the integral equations integrals in (\ref{NewcombO2}) and (\ref{Newcombphi2}) except $\phibarPOne$ are constants independent of $\alphaO$ as required. Therefore, we can set $\phibarOne = c\:\alphaO$ where $c$ is chosen so be a non-zero constant so that the constraints (\ref{NewcombO2}) and hence (\ref{Newcombphi2}) are satisfied.

\section{Physical interpretation of the free-functions}\label{App:AppendixB}
We shall now compare our approach with the well-known approach employing flux coordinates \citep{hudsonKrauss20173D_cont_B,helander2014theory}. In flux-coordinates $(\psi,\theta_f,\varphi_f)$, with the magnetic field given by $\B= \dl \psi \times \dl \theta_f + \iotabar(\psi)  \dl \varphi_f \times \dl \psi$, the magnetic differential equation takes the form
\begin{align}
\BD f= \frac{1}{\sqrt{g}}\lbr \del_{\varphi_f}+\iotabar\: \del_{\theta_f}\rbr f =G.
\label{MDEfluxcoord}
\end{align}
Here, $1/\sqrt{g}=\BD \varphi_f$ is the Jacobian and $\iotabar(\psi)$ is the rotation transform with the magnetic shear $\iotabar'\neq 0$. In Fourier space with $f=\sum_{(m,n)}f^{mn}(\psi)e^{i (m \theta_f -n \varphi_f)}$ etc. and a given $G$, we can write the general solution of the MDE \eqref{MDEfluxcoord} \citep{hudsonKrauss20173D_cont_B} as
\begin{align}
f^{mn} = \frac{i\lbr \sqrt{g}G\rbr}{\iotabar m -n}^{mn}+ \Delta^{mn} \delta(\iotabar m -n)
\label{MDEfourier}
\end{align}
with an undetermined constant $\Delta^{mn} $. The Delta-function term is a homogeneous solution of the MDE \eqref{MDEfluxcoord}. In real space, in the vicinity of the rational surface for which $\iotabar_0 = n/m$, the Delta-singular solution takes the form of a free-function $\bar{f}(\psi,\alphaO_f)$, with $\alphaO_f=\varphi_f - \iotabar_0  \theta_f$. We note here that just like the undetermined constant $\Delta^{mn}$, the functional form of $ \bar{f}$ can only be determined through boundary conditions and fluxes. The Delta-function term, although a singularity, is integrable and hence allowed. This is in contrast to the first term in \eqref{MDEfourier}, which clearly shows the ``small-divisor" problem and leads to an unphysical logarithmic singularity of $f$ in real-space near rational surfaces \citep{loizu2015jumpiota,hudsonKrauss20173D_cont_B}. 

We now briefly discuss the connection with our approach, which does not employ flux-coordinates. We expand near a low-order rational surface with a shear weak enough that the singularities on the low-order rational surfaces are well-separated. We use the fact that there are many functions $G$ for which there are good solutions and arrange the situation so that $G$ at each order is of the correct form to avoid the singular divisor problem. Our approach shows that we can use the $O(n-1)$ Delta-singular solutions (or free-functions) to eliminate the $(\iotabar m -n)^{-1}$ singularity occurring at $O(n)$.

In section \ref{sec:induct}, we have shown that at each order $n$, we encounter the MDEs \eqref{MDEalphaN} and \eqref{MDEphiNp1} that determine $\alphaN$ and $\phiNpOne$, and we inherit two free-functions $\phibarN(\alphaO)$ , $\alphabarNV{n-1}(\alphaO)$ from $O(n-1)$. Let us now discuss the physical quantities these free-functions are associated with.
From the definition of rotation transform and magnetic shear
\begin{align}
\iotabar = \frac{\BD \theta}{\BD \varphi} = -\frac{\alpha_{,\varphi}}{\alpha_{,\theta}}, \quad \iotabar_{,\psi}=\frac{-1}{\alpha_{,\theta}}\lbr\del_\varphi +\iotabar \del_\theta \rbr\alpha_{,\psi},
\end{align}
we find that the free-functions $\alphabarPOne,..\alphabarNmOneP$ represent averaged magnetic shear and its derivatives on the rational surface. 
Next, from the equations (\ref{currentJ},\ref{a3} and \ref{MDEzeta}) we find that the parallel current
\begin{align}
j_\parallel = -\zeta_{,\alpha}= \del_\alpha \phi_{,\psi} +..
\label{parallelJ}
\end{align}
Therefore, the free-functions $\phibarPOne(\alphaO),..\phibarNP(\alphaO)$ are related to the parallel DC current and its derivatives on the lowest order rational surface.

As shown in \eqref{NewcombalpN}, upon averaging the MDEs obtained at $O(n)$ over the closed field line on the rational surface, we obtain several resonant terms. Some of the resonant terms occur due to the free-functions, whereas others occur due to the nonlinear beating of non-resonant terms.  The net sum of all such resonant terms must vanish to prevent the logarithmic singularities. Provided \eqref{determ} is satisfied, we can use the free-functions to eliminate the resonances. The lowest order $n=0$, discussed in Appendix \ref{App:AppendixA}, needs special care because there are three integral constraints, (\ref{NewcombR1}),(\ref{NewcombO2}) and (\ref{Newcombphi2}) and only one free-function $\phibarOne(\alphaO)$. Thus, we have demonstrated how we can obtain a so-called ``healed-configuration'' by adjusting the shear and the parallel current profiles. In ideal MHD only two profiles can be specified independently. In axisymmetry, the plasma boundary is arbitrary. However, in full three-dimensions, the removal of the resonant terms using the shear and parallel current profile shows that the plasma boundary cannot be arbitrary and must be self-consistent with the solution. The final solutions for the magnetic field, pressure and rotation transform are all smooth in this low-shear regime.

\bibliographystyle{jpp}

\bibliography{plasmalit}

\end{document}